\documentclass[reprint,preprintnumbers,aps,showpacs,citeautoscript]{revtex4-1}

\usepackage[utf8]{inputenc}
\usepackage{amsmath}
\usepackage{graphicx}
\usepackage{amssymb}
\usepackage{times}
\usepackage{color}
\usepackage{float}
\usepackage{multirow}
\usepackage{subcaption}
\usepackage{braket}

\begin{document}

\title{Theoretical polarization of zero phonon lines in point defects}

\author{Joel Davidsson}
\email{joel.davidsson@liu.se}
\affiliation{Department of Physics, Chemistry and Biology, Link\"oping University, Link\"oping, Sweden}

\begin{abstract}

In quantum technologies, point defects in semiconductors are becoming more significant. Understanding the frequency, intensity, and polarization of the zero phonon line is important. The last two properties are the subject of this paper. I present a method for calculating these properties and show the importance of using wave functions from both the ground and excited state. The validity of this method is demonstrated on the divacancy in 4H-SiC. Here, the calculated polarization and lifetimes are in excellent agreement with experimental measurements. In general, this method can help to identify point defects and estimate suitable applications.

\end{abstract}

\maketitle

%%%%%%%%%%%%%%%%%%%%%%%%%%%%%%%%%%%%%%%%%%%%%%%%%%%%%%%%%%%%%%%%%%%%%%%%%%%%%%%%%%%%%%%%%%%%%%%%%%%%%%%%
%%%%%%%%%%%%%%%%%%%%%%%%%%%%%%%%%%%%%%%%%%%%%%%%%%%%%%%%%%%%%%%%%%%%%%%%%%%%%%%%%%%%%%%%%%%%%%%%%%%%%%%%

\section{Introduction}
Applications for point defects in semiconductors are many and include qubits~\cite{Jelezko:PSS2006,Hanson:Nature2008,Awschalom2013}, quantum sensing applications~\cite{Kucsko2013,Falk2014,Balasubramanian:Nature2008}, and single photon emitters~\cite{Childress:PRL2006,Aharonovich:NL2009,Kolesov2012,NatMat14,Aharonovich2016}.
Studying point defects is an active field and for the applications mentioned, properties like isolated spin, long coherence time, and room temperature operations~\cite{Balasubramanian:NatMat2009,Christle2014,Widmann2014} are desired.
Also, the possibility to integrate point defects in quantum technologies and semiconductor devices is preferred.
Before these applications can be realized, the point defects present in the material of interest need to be identified.

To achieve accurate identification of a point defect, many different properties such as zero phonon line (ZPL), zero field splitting, and hyperfine coupling parameters need to be compared~\cite{methodology_paper}.
This approach works best for high spin and symmetry point defects since zero field splitting and hyperfine coupling parameters only exist for high spin defects.
However, even if these properties formally should exist in high spin defect, the low symmetry configuration can make it difficult to measure a clear signal in electron spin resonance experiment, see the missing hyperfine for low symmetry configurations in Table~III in Ref.~\onlinecite{methodology_paper}.
One of these low symmetry configurations is a promising candidate for exploring spin and optical dynamics~\cite{miao2019electrically}.
For both low spin and symmetry point defects, the polarization of the ZPL is a great addition to the properties used to understand the point defect. 
It is easy to measure regardless of the defect spin.
For low spin defects, ZPL and the corresponding polarization may be the only measurements available.
For low symmetry defects, group theory can not determine the polarization, hence calculations are required.
For applications, like single photon emitters, a ZPL with a large intensity is required~\cite{udvarhelyi2019spectrally}.
To determine if point defect will be a bright single photon emitter, an accurate ZPL intensity is needed.
The intensity will, in turn, affect the lifetime of the excited state.
The effect of ion relaxation and corresponding electronic structure change of the excited state has on the polarization and intensity of ZPL has not yet been determined.
This paper aims to show how these changes of the excited state affect both the polarization and intensity of ZPL.

In this paper, the divacancy in 4H-SiC is used to demonstrate the validity of this method.
For material with different polytypes, one point defect can have many configurations.
For instance, in 4H-SiC, the divacancy has four different configurations.
In experiments, divacancy related luminescence lines are named UD-2 group~\cite{magnusson2005} and PL1-4 lines~\cite{Koehl11,Falk2013}.
These configurations have previously been identified by comparing theoretical calculations with experimental measurements~\cite{methodology_paper}.
The different divacancy configurations in 4H-SiC exist due to the different non-equivalent symmetric sites for the Si and C atoms.
4H-SiC consists of four layers: two layers, which have a hexagonal-like environment denoted with $h$; and two layers, which have a cubic-like environment denoted with $k$.
In the case of the divacancy, the following notation V$_{\text{Si}}$-V$_{\text{C}}$ denotes which layers the point defect belongs to.

% all defects are in the yz-plane
\begin{figure}[h!]
   \includegraphics[width=0.9\columnwidth]{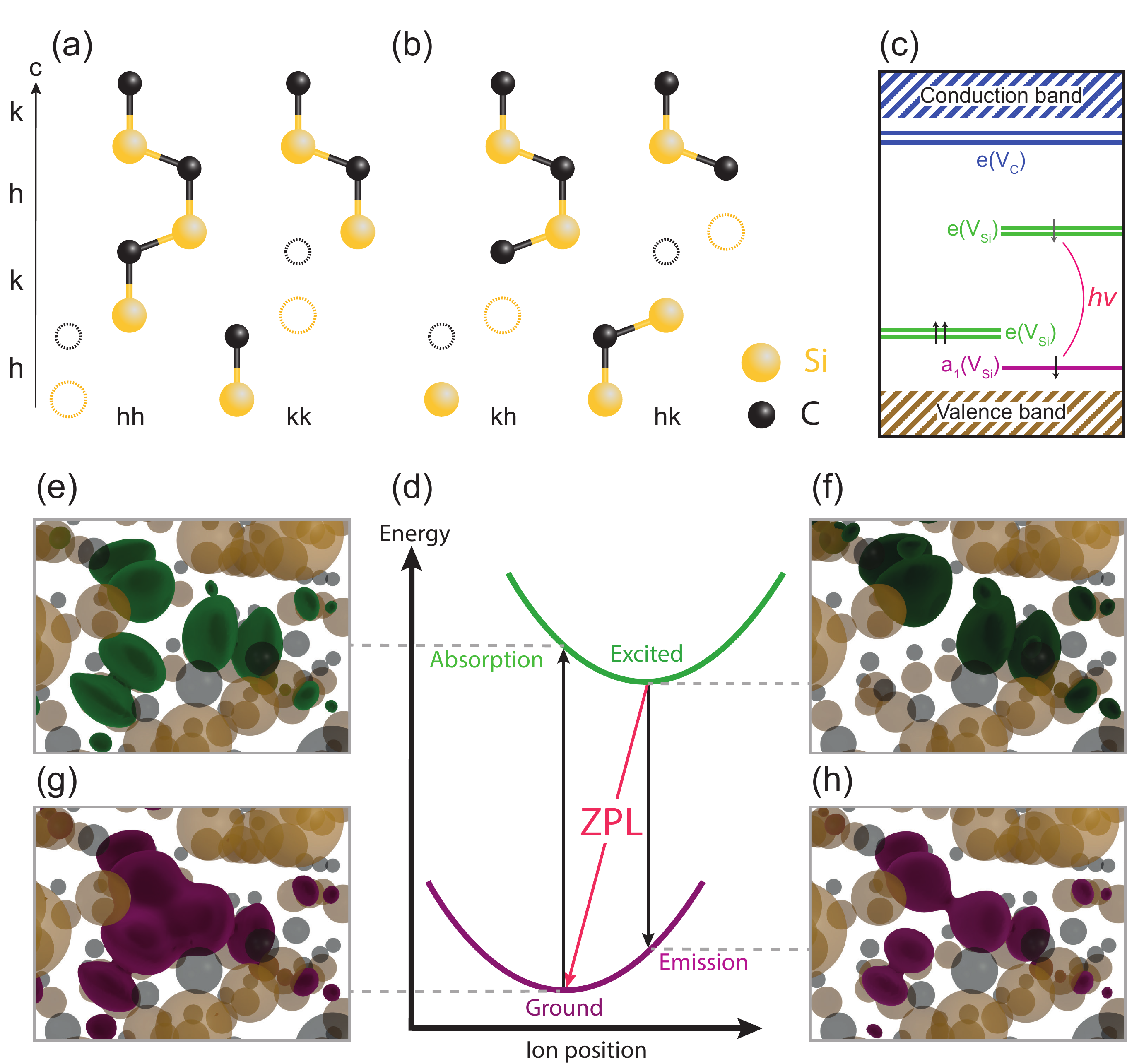}
	\caption{(a) and (b) shows the $C_{3v}$ and $C_{1h}$ point group symmetry divacancy configurations. (c) shows the schematic electronic structure for the divacancy. (d) shows the excitation cycle. Where, (e)-(h) shows the partial densities of the (purple) $a_1$ and (green) $e$ orbitals at the different excitation stages for the $hh$ configuration.}
	\label{fig:intro} 
\end{figure}

Figure~\ref{fig:intro}(a) shows the $C_{3v}$ point group symmetry configurations, $hh$ and $kk$.
Figure~\ref{fig:intro}(b) shows the $C_{1h}$ point group symmetry configurations, $kh$ and $hk$, which can exist in three different orientations due to the symmetry of the defect-free crystal.
The polarization of the different configuration follows the point group symmetries.
For the $C_{3v}$ point group symmetry, the polarization must be in the basal plane (x- and y-direction), since no change takes place in the z-direction.
However, the $C_{1h}$ point group symmetry can have polarization in any direction.
Figure~\ref{fig:intro}(c) shows a schematic diagram of the electronic structure for the divacancy.
Here, the transition between the defect orbitals in the band gap is highlighted.
In the case of the divacancy in 4H-SiC, this transition between the $^3\mathrm{E}$ and $^3\mathrm{A}_2$ many-body state produces the ZPL line.
Figure~\ref{fig:intro}(d) shows the ZPL line as the transition between the excited state including the ion relaxation and ground state.
Figure~\ref{fig:intro}(e)-(h) shows the k-point averaged partial densities for the $a_1$ and $e$ orbitals during each step in the excitation cycle for the $hh$ configuration.

In this paper, I present how to calculate and converge the ZPL polarization.
The method consists of calculating the transition dipole moment (TDM) between Kohn-Sham (KS) orbitals from the ground and excited wave functions.
Using only one wave function would give the polarization of the absorption, not of the ZPL.
These polarizations differ due to the ionic relaxation and corresponding electronic structure change of the excited state.
This method is demonstrated on the divacancy in 4H-SiC, where the ion relaxation and corresponding electronic structure change is necessary to reproduce the polarization of the low symmetry configurations as well as the lifetimes of all configurations.

\section{Method}
The TDM $\boldsymbol{\mu}$ from initial orbital $\psi_i$ to final orbital $\psi_f$ is defined as 
\begin{equation}
\boldsymbol{\mu}=\braket{\psi_f|q\mathbf{r}|\psi_i}.
\end{equation}
This equation is rewritten into reciprocal space, which for a periodic system would give 
\begin{equation}
\boldsymbol{\mu}_k=\frac{i \hbar}{(\epsilon_{f,k}-\epsilon_{i,k})m}\braket{\psi_{f,k}|\mathbf{p}|\psi_{i,k}}.
\label{eq:tdm}
\end{equation}
Where $\epsilon_{f,k}$ and $\epsilon_{i,k}$ are the eigenvalues of the final and initial orbital.
The subscript k is added to emphasize the k-point dependency of the TDM (cf. Figure~1(e) in Ref.~\onlinecite{jiang2018role}).
Usually, the initial and final orbitals are two orbitals in the same wave function.
However, the ZPL is a transition between two different electronic structures thus two wave functions are needed.
In this article, the final orbital is taken from the excited wave function while the initial orbital is taken from the ground wave function.

The ground and excited wave functions are obtained using density functional theory~\cite{Hohenberg64,Kohn65,ivady2018first} (DFT) and the constrained occupation DFT method~\cite{Gali:PRL2009}, respectively.
In practice, I use the Vienna Ab initio Simulation Package (VASP)~\cite{VASP,VASP2}, which is based on projector augmented wave (PAW) method~\cite{PAW,Kresse99}.
The calculations use the semi-local functional of Perdew, Burke, and Ernzerhof (PBE)~\cite{PBE} to describe the exchange-correlation.
The plane wave (kinetic) energy cutoff is set to 600 (900) eV.
The energy criterion for the electronic self-consistent cycle is set to $10^{-6}$ eV and the ion relaxation to $5 \times 10^{-5}$ eV.
To avoid wrap-around errors, the grid for the Fast Fourier transform is set to twice the largest wave vector.
The k-point grid is constructed using a Monkhorst-Pack~\cite{monkhorst1976special} grid.
The Brillouin zone was sampled using the tetrahedron method, except for the $\Gamma$-point calculations where Fermi smearing is used, with smearing width of 1 meV.
The supercell of 4H-SiC contains 576 atoms, which is the unit cell repeated (6, 6, 2) and thus follow the symmetry of the crystal.
The only symmetry used in the calculations is $\Psi_k=\Psi_{-k}^*$.

After the density has reached self-consistency, the ground and excited wave functions are post-processed using PyVaspwfc~\cite{zheng_2019}, \emph{a python class for VASP WAVECAR}.
This code reads the plane wave coefficient of the pseudo wave function and calculates the TDM between different KS orbitals according to Eq.~\eqref{eq:tdm}, I modified this code to handle reading two wave functions instead of one.
When using the pseudo wave functions, the contribution inside the PAW spheres is neglected.
This contribution is about 5\% of the total wave function~\cite{Ivady2014} and should only affect the intensity of the TDM since the polarization is determined mostly from the interstitial region.

In general, each component in the TDM vector is a complex number.
The amplitude of these complex numbers gives the polarization, hence the phase can be neglected.
The absolute value of TDM is obtained by averaging over the k-points,
\begin{equation}
\bar{\boldsymbol{\mu}}=\sum_k w_k (|\mu_{kx} | \boldsymbol{\hat{x}} + |\mu_{ky} | \boldsymbol{\hat{y}} + |\mu_{kz} | \boldsymbol{\hat{z}}) \bigg/ \sum_k w_k,
\end{equation}
where $w_k$ is the weight of the k-point.
The average TDM vector $\bar{\boldsymbol{\mu}}$ shows the polarization--the amplitude of this vector can be used to estimate the spontaneous transition rate between the excited and ground state.
This is calculated using the Einstein coefficient $A$ which is defined as
\begin{equation}
A=\frac{n (2\pi)^3 \nu^3 |\bar{\boldsymbol{\mu}}|^2}{3 \epsilon_0 h c^3}.
\label{eq:lifetime}
\end{equation}
This equation contains the following constants: $\nu$ is the transition frequency of the ZPL; $\epsilon_0$ is the vacuum permittivity; $h$ is the Planck constant; $c$ is the speed of light; and $n$ is the refractive index, which for 4H-SiC is $n=2.6473$.
The inverse of A is the radiative lifetime $\tau$ between the excited and ground state.

\section{Results}

Here follow the results from testing the method on the different configurations of the divacancy in 4H-SiC.
For this defect, $e$ is the final orbital and $a_1$ is the initial orbital in Eq.~\eqref{eq:tdm}.
In the excited state, the degenerate $e$ orbital is singly occupied in one spin channel leading to a Jahn-Teller splitting.
For the relaxation of the excited state, the static Jahn-Teller is included while the dynamic Jahn-Teller is neglected.

\begin{figure}[h!]	\includegraphics[width=\columnwidth]{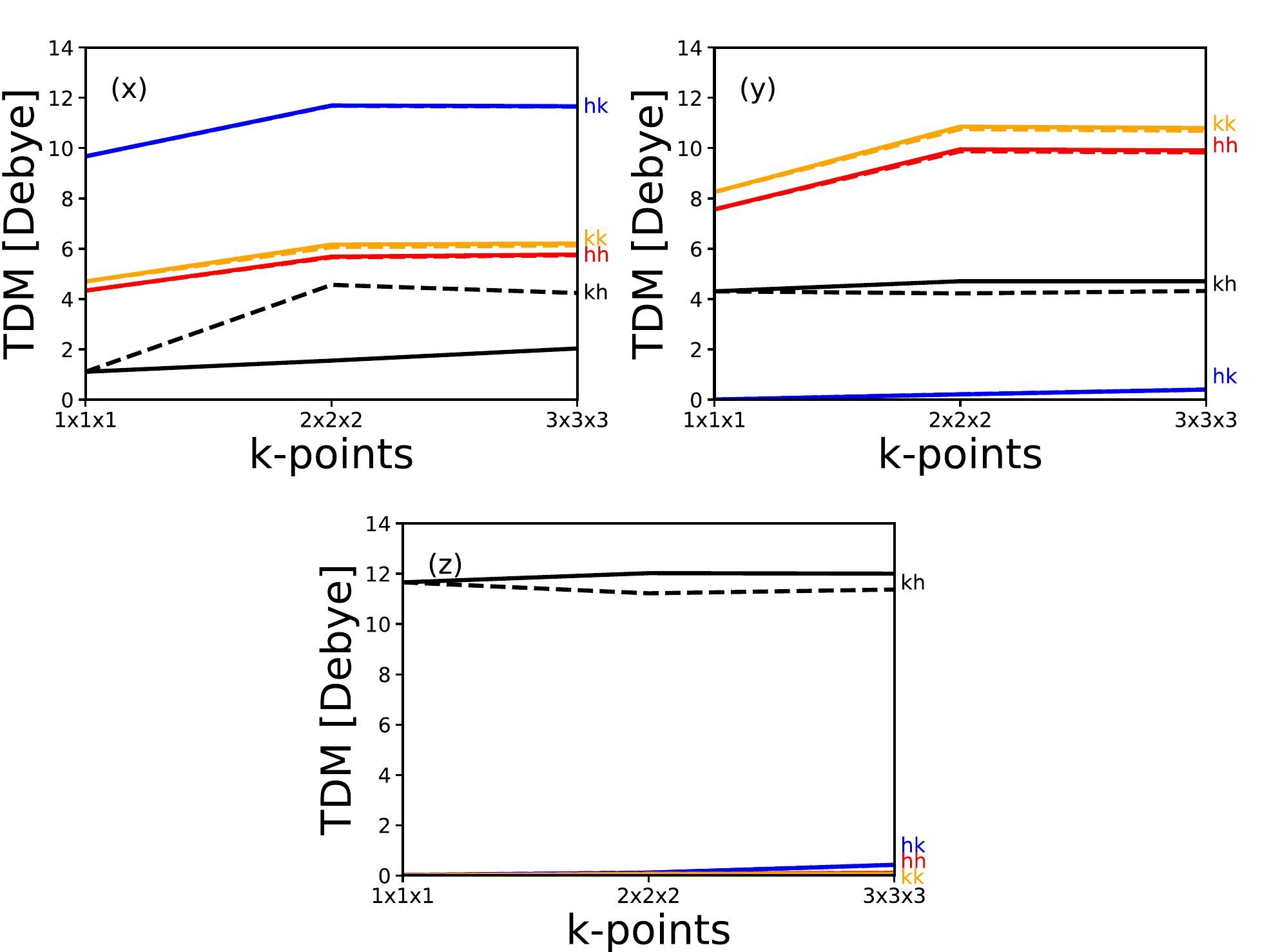}
	\caption{k-point convergence for $\bar{\boldsymbol{\mu}}$ for the $hh$, $hk$, $kh$, and $kk$ configurations. (x), (y), and (z) depicts the different directions of the $\bar{\boldsymbol{\mu}}$. In the z-direction, only the $kh$ is non-zero. The solid lines show the results using the $\Gamma$ point relaxed structure. The dashed lines show the results including ion relaxation for each k-point set.} 
	\label{fig:kpoints} 
\end{figure}

Figure~\ref{fig:kpoints} shows $\bar{\boldsymbol{\mu}}$ with changing k-point sampling.
First, convergence is reached for $2 \times 2 \times 2$ k-point set, which is a total of 8 k-points.
This is the same convergence as needed for the ZPL energy (cf. Figure~3(d) in Ref.~\onlinecite{methodology_paper}).
Second, the sampling of the Brillouin zone accounts for the main change in convergence.
Including relaxation for each k-point set only affects the $kh$ configuration.
This change is most visible in the x-direction, where including the relaxation for the $2 \times 2 \times 2$ k-point set gives a value of 4.6 Debye, whereas using the $\Gamma$-point relaxed geometry gives a value of 1.6 Debye.

As mentioned in the introduction, the high symmetry configurations ($hh$ and $kk$) can only be oriented in one way.
However, the low symmetry configurations ($kh$ and $hk$) exist in three orientations, each with a different $\bar{\boldsymbol{\mu}}$.
Here, the z-component, which is parallel to the c-axis of the supercell, is the same.
In the basal plane, the x- and y-components vary.
This change is a rotation of the $\bar{\boldsymbol{\mu}}$, therefore the sum of these components is the same.
Hence, the combined x- and y-components are called the perpendicular component ($\perp$\textbf{c}) and the z-component is called the parallel component ($\parallel$\textbf{c}).

\begin{figure}[h!]\includegraphics[width=\columnwidth]{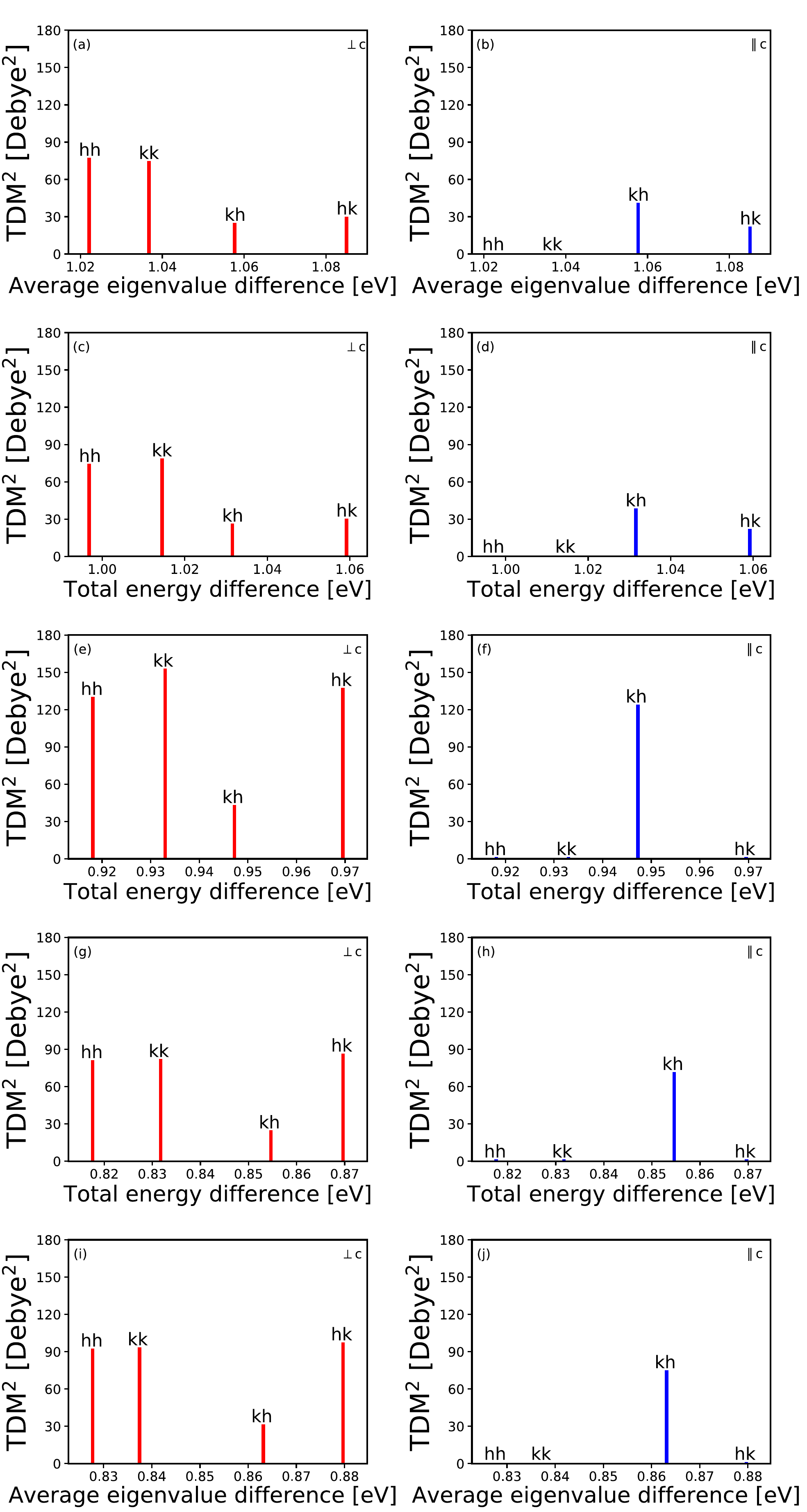}
	\caption{Theoretical polarization for absorption, ZPL, and emission. The left column shows the perpendicular components and the right column shows the parallel components. The absorption results are shown using $\bar{\boldsymbol{\mu}}^{gg,gg}$ in (a)-(b) and $\bar{\boldsymbol{\mu}}^{eg,gg}$ in (c)-(d). (e)-(f) shows the ZPL results $\bar{\boldsymbol{\mu}}^{ee,gg}$. The emission results are shown using $\bar{\boldsymbol{\mu}}^{ee,ge}$ in (g)-(h) and $\bar{\boldsymbol{\mu}}^{ge,ge}$ in (i)-(j). When the same wave function is used, (a)-(b) and (i)-(j), the x-axis is the average difference between the eigenvalues. Otherwise, the x-axis is the total energy difference.}
	\label{fig:main}
\end{figure}
% say something about the fig3 ab,ij?

Using the convergent $2 \times 2 \times 2$ k-point set with relaxation, $\bar{\boldsymbol{\mu}}$ is calculated for the absorption, ZPL, and emission in five different ways to show the effect of each change.
To keep track of the orbitals' origin, a superscript is added to $\bar{\boldsymbol{\mu}}^{ff,ii}$ to denote which wave function and geometry is used.
Index $ff$ and $ii$ indicate where the final and initial orbital originates from.
The first and second letter denotes the wave function and relaxed geometry, respectively, with the notation: $g$ for ground and $e$ for excited.
For the absorption, $\bar{\boldsymbol{\mu}}^{gg,gg}$ is calculated between the orbitals in only the ground wave function, whereas $\bar{\boldsymbol{\mu}}^{eg,gg}$ is calculated using the same geometry but the final orbital is taken from the excited wave function.
The ZPL is calculated between the final orbital from the excited wave function on the corresponding relaxed geometry and the initial orbital from the ground wave function on the corresponding relaxed geometry thus denoted $\bar{\boldsymbol{\mu}}^{ee,gg}$.
For the $hh$ configuration, Figure~\ref{fig:intro}(f) and (g) shows the orbitals used in $\bar{\boldsymbol{\mu}}^{ee,gg}$.
For the emission, $\bar{\boldsymbol{\mu}}^{ee,ge}$ is calculated between the ground and excited wave functions on the excited geometry while $\bar{\boldsymbol{\mu}}^{ge,ge}$ is calculated with only ground wave function on the excited geometry.
The results for the different transitions are presented in Figure~\ref{fig:main}.
Here, no apparent difference in the absorption and emission polarization is observed when going from only ground wave function to both ground and excited wave functions while keeping the geometry fixed.
The electronic excitation is not enough to change the polarization.
Only when including the ion relaxation and corresponding electronic structure change does both the polarization and intensities change.
For the ZPL spectra, the intensities roughly double compared with both absorption and emission.
The $hk$ configuration changes from both parallel and perpendicular polarization in the absorption spectra to only perpendicular polarization in the ZPL spectra and stays that way also in the emission spectra.
The choice of functional when calculating the ZPL polarization is marginal, see Appendix~\ref{app:func_dep} for further discussion.

\begin{figure}[h!]	\includegraphics[width=\columnwidth]{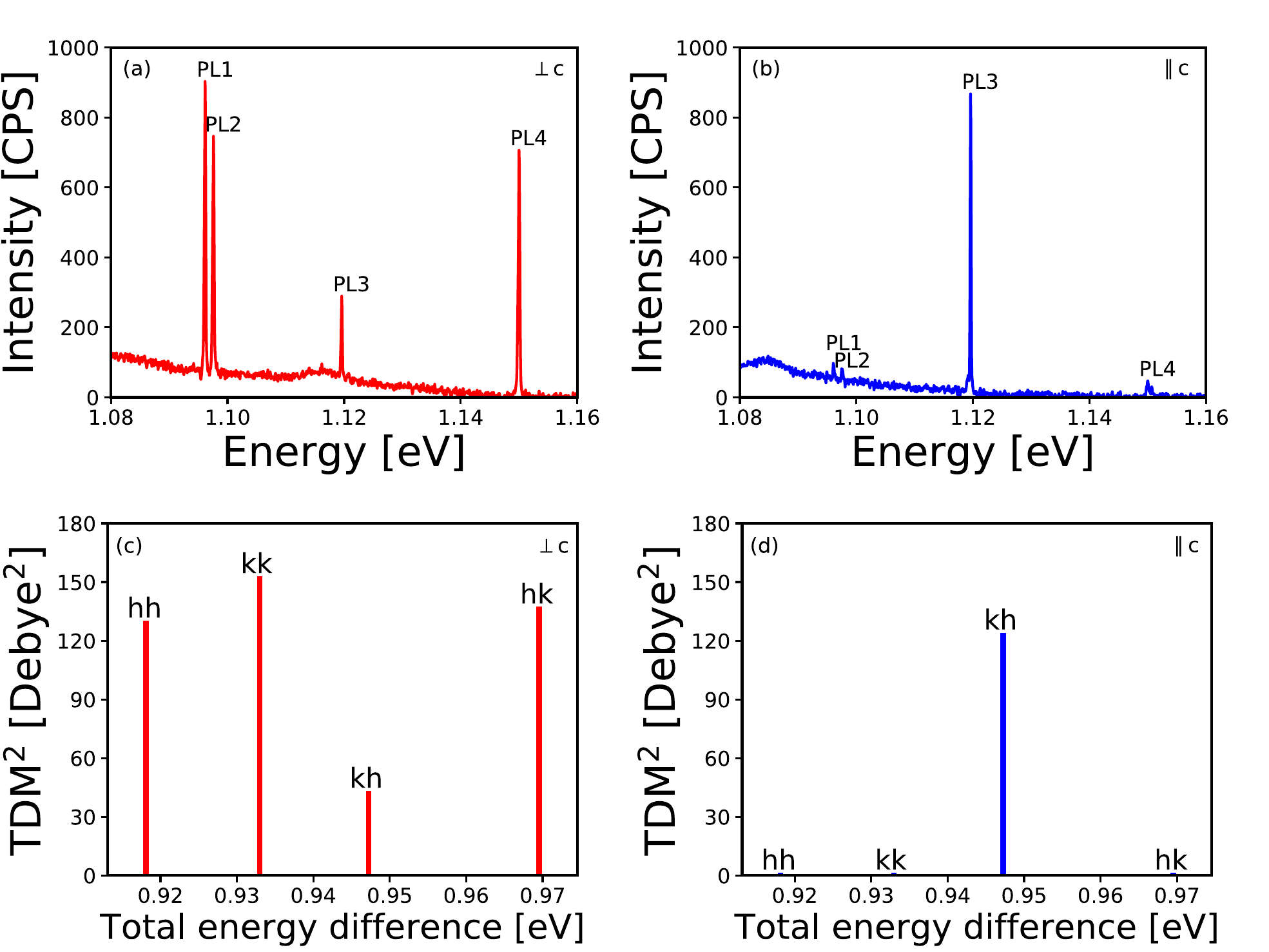}
	\caption{Polarization data for the divacancy in 4H-SiC. The top row is the experimental data, which was part of a series of photoluminescence measurements~\cite{gallstrom2015optical}, where (a) and (b) shows the perpendicular and parallel components, respectively. The bottom row is the theoretical results, (c) shows perpendicular and (d) parallel components of $\bar{\boldsymbol{\mu}}^{ee,gg}$.} 
	\label{fig:zpl_exp} 
\end{figure}

In Figure~\ref{fig:zpl_exp}, the theoretical polarization results are compared to the experimental measurements~\cite{gallstrom2015optical}.
Here, the ZPL $\bar{\boldsymbol{\mu}}^{ee,gg}$ reproduce the trend present in experiment.
In the perpendicular direction, the experimental data is large for PL1, PL2, and PL4 but small for PL3, see Figure~\ref{fig:zpl_exp}(a).
The theoretical results reproduce this trend, see Figure~\ref{fig:zpl_exp}(c).
Here, the intensity has been reversed for PL1 and PL2 when compared to experiment.
In the parallel direction, only PL3 is visible for experiment data and $kh$ for theoretical results, see Figure~\ref{fig:zpl_exp}(b) and (d), respectively.

The radiative lifetimes for the different configurations are calculated from the $\bar{\boldsymbol{\mu}}$ intensities using Eq.~\eqref{eq:lifetime}.
For the absorption, using $\bar{\boldsymbol{\mu}}^{gg,gg}$ or $\bar{\boldsymbol{\mu}}^{eg,gg}$ , the lifetime ranges from 38 to 50 ns ($kk$: 38.49 ns, $hh$: 39.05 ns, $kh$: 42.39 ns, and $hk$: 50.82 ns).
For the ZPL, using $\bar{\boldsymbol{\mu}}^{ee,gg}$ , the lifetime ranges from 16 to 23 ns ($kh$: 16.41 ns, $hk$: 18.52 ns, $kk$: 18.65 ns, and $hh$: 23.01 ns).
The ZPL lifetimes are shorter than the absorption due to the larger $\bar{\boldsymbol{\mu}}$.
Note, that the low symmetry configurations have the shortest lifetime for the ZPL whereas it is the high symmetry configurations for the absorption.

\section{Discussion}
It is clear from Figure~\ref{fig:kpoints}, $2 \times 2 \times 2$ k-point grid is needed to converge the TDM results.
The high symmetry configurations converge fastest, with no change going to $3 \times 3 \times 3$.
The low symmetry configurations have a small increase going to $3 \times 3 \times 3$, most noticeable for the $kh$ in the x-direction and $hk$ in the y-direction.
Also, including relaxation for the $2 \times 2 \times 2$ grid had a large effect in the $kh$.

Figure~\ref{fig:main} shows the differences in polarization and intensities between using ground and excited wave functions.
First, only going from ground wave function to ground and excited wave functions do not affect the polarization or intensities as seen for both absorption and emission.
The electronic change of the wave function is not enough to change the intensities because the single orbitals themselves barely change during the excitation.
Only when the ions in the excited state are allowed to relax and the single orbitals change, do the polarization and intensities change.
The orbital change due to the ion relaxation can be seen in Figure~\ref{fig:intro} going from (e) to (f).
The higher intensities indicate a higher probability of optically relaxing back to the ground state.
In the case of the divacancy, this amplitude roughly doubles compared with absorption.
For a general point defect, it may change orders of magnitude depending on how large the ion relaxation and orbital change in the excited state is.

For the polarization, the high symmetry configurations ($hh$ and $kk$) do not change their polarization regardless of absorption, ZPL, or emission.
As discussed in the introduction, the polarization must follow the point group symmetry, which states that the change between $a_1$ and $e$ orbital in $C_{3v}$ symmetry is zero in the parallel direction, as can be seen in Figure~\ref{fig:main}.
The amplitude of the perpendicular components can still change, which it does between the absorption and ZPL.
For the low symmetry configurations ($kh$ and $hk$), any polarization is possible from a group symmetry point of view.
This can also be seen in Figure~\ref{fig:main}, the polarization changes quite drastically for these configurations when going from absorption to ZPL.
When these results are compared to experiment as in Figure~\ref{fig:zpl_exp}, it is clear that the change in electronic structure due to ion relaxation of the excited state is needed to correctly reproduce the polarization.
This is most noticeable in the parallel direction for the $hk$ configuration which is only zero if two wave functions are used, see Figure~\ref{fig:zpl_exp}(d).

The theoretical results can be used to interpret the experimental data.
The ZPL energy and the limitations of the PBE functional have on the results has been discussed in detail in Ref.~\onlinecite{methodology_paper}.
When comparing the theoretical polarization to experiment, the intensity of the ZPL line is determined by the TDM.
However, when comparing with experiment, other factors affect the measured intensity.
These include the concentration of the point defects, non-radiative relaxation, and vibrations.
I assume the concentration of different point defects is the same.
Any differences in concentration should only have a minor effect on the intensity.
The percentage of non-radiative relaxation is also assumed to be the same for the different configurations as well as the vibrational effects in both the ground and excited state.
Calculating any differences in these rates and effects would require time-consuming ab-initio calculations.
In this article, the vibrational effects and zero-point motion has been assumed to be the same in both the ground and excited state.
This does not have to be the case as argued in Ref.~\onlinecite{alkauskas2014first} but as also mentioned in this reference, including these effects are highly non-trivial.
Overall, these effects should only affect the intensities of the TDM, not the polarization.

Here follows how to identify the different configurations for the divacancy in 4H-SiC using the ZPL energy and polarization results.
First, the PL3 line clearly is the $kh$ configuration.
This conclusion is reached by comparing the polarization results presented in Figure~\ref{fig:zpl_exp}(b) and (d).
When comparing Figure~\ref{fig:zpl_exp}(a) and (c), PL4 can be identified with $hk$.
Here, both the polarization and ZPL energy provide a clear identification.
For the PL1 and PL2, the relative difference between the intensities and ZPL energies are too low to accurately identify the configurations.
As discussed in the previous paragraph, small changes could be due to differences in concentration, non-radiative relaxation, and vibration and would require further time-consuming ab-initio calculations.
An indication that the vibration could be important can be seen if the temperature is increased (20 K), then the intensities reverse for the PL1 and PL2 as shown in Figure~1b in Ref.\onlinecite{Koehl11}.
Alternatively, additional properties such as hyperfine and zero-field splitting, can be used to make an accurate identification for these configurations~\cite{methodology_paper}.
When using only the ZPL energy and polarization results, the low symmetry configurations can be accurately identified, where PL3 is $kh$ and PL4 is $hk$, which strengthen the previous identification~\cite{methodology_paper} where hyperfine for these configurations are missing. 

The radiative lifetimes calculated in this article show that the ZPL lifetimes are shorter (16 to 23 ns) than the absorption lifetimes (38 to 50 ns).
Previous theoretical calculation (configuration interaction approach with constrained random phase approximation) obtains a lifetime of 63 ns for the $hh$ configuration~\cite{bockstedte2018ab}.
This lifetime is in the same range as the absorption lifetimes presented in this paper showing that including a better description of the correlated states does not improve the lifetime results.
See Appendix~\ref{app:func_dep} for additional discussion about the choice of functional and the effect on the lifetime.
However, including the ion relaxation of the excited state and corresponding electronic structure change produces results closer to experiment. 
The ZPL lifetimes are closer to experiement but still slightly overestimate the experimental range of 12 to 15 ns~\cite{Falk2014}.
However, these lifetimes were measured at 20 K, which may speed up the radiative decay.
Another source of error for the ZPL lifetimes is that the contributions inside the PAW spheres are neglected.
These contributions may slightly increase the amplitude of the $\bar{\boldsymbol{\mu}}$, which in turn would lower the lifetime further.
Overall, the ZPL lifetimes are still closer to experiment than the absorption lifetimes.
Further supporting the importance of including electronic structure change due to the ion relaxation of the excited state when calculating TDM.
Additional experimental lifetimes at 0 K are needed for a better comparison with the lifetimes presented in this article.

\section{Conclusion}
To conclude, this work shows how to obtain the theoretical ZPL polarization and intensity as well as the importance of including electronic structure change due to the ion relaxation of the excited state.
For the ZPL, the most important part is using the best physical representation of both the excited and ground state.
Before the excited electron returns to the ground state, the best physical description of that electron is provided by the excited wave function and corresponding geometry.
The method calculates the transition dipole moment between Kohn-Sham (KS) orbitals from the ground and excited wave functions.
This method not only provides the polarization, which could be used as a compliment when identifying the different configurations, but also the total intensity, which could be used to estimate radiative lifetimes.
For the divacancy in 4H-SiC, the calculated polarization and the estimated lifetimes are in excellent agreement with experimental measurements.
Here, the low symmetry configurations are identified by comparing the polarization results and experimental measurements.
In general, this approach may be especially useful for identifying low spin and symmetry configurations.
%This method is needed to describe the ZPL for defects with a large ion relaxation in either the ground or excited state, such as defects with Jahn-Teller effect, in soft materials, or in materials with light elements.
The ability to calculate polarization, intensities, and radiative lifetimes is a great addition to the analysis of any point defect.
With these properties, one could screen point defects for their potential use in applications, like single photon emitters where large intensities and short lifetimes are required.

\appendix
\section{Functional dependence on polarization and lifetime}
\label{app:func_dep}

To investigate the effect of the functional, the TDM is also calculated with the screened hybrid functional of Heyd, Scuseria, and Ernzerhof (HSE06)\cite{HSE03,HSE06}.
The hybrid functional is calculated on the relaxed PBE functional geometry, with $\Gamma$-point only, for both the ground and excited state with corresponding PBE wave function as a starting point.
The electronic self-consistent cycle is reduced to $10^{-4}$ eV and no ion relaxation takes place due to the high computational load.
Only small changes are expected if a larger k-points mesh is used or ion relaxation is included for the HSE calculations.
Table~\ref{tab:pbe_hse} shows the TDM results for absorption and ZPL for the two functionals.

\begin{table}[h!]
\caption{$\bar{\boldsymbol{\mu}}$ for each divacancy configuration for absorption ($\bar{\boldsymbol{\mu}}^{gg,gg}$) and ZPL ($\bar{\boldsymbol{\mu}}^{ee,gg}$) with PBE and HSE functional.}
\vspace{-3mm}
\begin{ruledtabular}
\begin{tabular} {c|c||cc|cc}
& & \multicolumn{2}{c|}{Absorption}  & \multicolumn{2}{c}{ZPL}  \\
Configuration & TDM & PBE  & HSE & PBE & HSE \\
\hline
\multirow{4}{*}{$hh$} & $\bar{\mu}_x$ & 3.16 & 2.61 & 4.34 & 4.48   \\ 
& $\bar{\mu}_y$ & 5.27 & 4.22 & 7.57 & 7.79  \\ 
& $\bar{\mu}_z$ & 4.58$\cdot10^{-7}$ & 2.88$\cdot10^{-6}$ & 7.94e$\cdot10^{-3}$ & 5.26$\cdot10^{-3}$  \\
& $|\bar{\boldsymbol{\mu}}|^2$ & 37.74 & 24.62 & 76.04 & 80.69   \\ 
\hline
\multirow{4}{*}{$kk$} & $\bar{\mu}_x$ & 2.43$\cdot10^{-3}$ & 1.86$\cdot10^{-2}$ & 4.70 & 4.92   \\ 
& $\bar{\mu}_y$ & 6.09 & 5.00 & 8.26 & 8.59  \\ 
& $\bar{\mu}_z$ & 3.54$\cdot10^{-4}$ & 3.34$\cdot10^{-4}$ & 4.73$\cdot10^{-3}$ & 7.68$\cdot10^{-3}$  \\
& $|\bar{\boldsymbol{\mu}}|^2$ & 37.10 & 25.03 & 90.26 & 97.96   \\ 
\hline
\multirow{4}{*}{$hk$} & $\bar{\mu}_x$ & 9.06$\cdot10^{-5}$ & 5.43 & 9.68 & 9.62   \\ 
& $\bar{\mu}_y$ & 2.62 & 7.67$\cdot10^{-6}$ & 3.31$\cdot10^{-3}$ & 3.42$\cdot10^{-3}$  \\ 
& $\bar{\mu}_z$ & 8.70 & 2.16$\cdot10^{-5}$ & 1.75$\cdot10^{-2}$ & 1.29$\cdot10^{-2}$  \\
& $|\bar{\boldsymbol{\mu}}|^2$ & 82.61 & 29.43 & 93.61 & 92.48   \\ 
\hline
\multirow{4}{*}{$kh$} & $\bar{\mu}_x$ & 1.18$\cdot10^{-5}$ & 7.13$\cdot10^{-5}$ & 1.11 & 2.35   \\ 
& $\bar{\mu}_y$ & 2.96 & 2.33 & 4.31 & 3.98  \\ 
& $\bar{\mu}_z$ & 8.51 & 6.37 & 11.66 & 10.40  \\
& $|\bar{\boldsymbol{\mu}}|^2$ & 81.10 & 46.01 & 155.64 & 129.60   \\ 
\end{tabular}
\end{ruledtabular}
\label{tab:pbe_hse}
\vspace{-2mm}
\end{table}

For the absorption results, the intensity goes down for all configurations when using the HSE functional.
The largest change is for the $hk$ where the polarization also changes to only x-direction.
For the absorption, the HSE functional has a large effect on the electronic structure thus providing the most accurate results.
However, for the ZPL results, there is only a small intensity difference between the PBE and HSE results.
For the high symmetry configurations the intensity increases with the HSE functional while for the low symmetry configurations, it decreases.
This indicates that a more accurate exchange-correlation functional is not necessarily needed to describe the polarization and intensity between the excited and ground state.
This is most likely because the TDM is an electrostatic property and the orbital change of the excited state is greater than the change due to exchange-correlation effects.
Hence, the main contribution comes from the use of different wave functions to describe the physical states when calculating the TDM for the ZPL.

There is only a small change for the TDM results when switching from PBE to HSE functional, what about the lifetime.
The radiative lifetimes depend on both the TDM and energy of the ZPL, see Eq.~\eqref{eq:lifetime}.
It is well known that the PBE ZPL energies are underestimated and the HSE ZPL energies are accurate compared to experiment~\cite{methodology_paper}..
For the absorption ($\bar{\boldsymbol{\mu}}^{gg,gg}$) the HSE lifetime ranges from 38 to 75 ns ($kh$: 38.07 ns, $hk$: 56.55 ns, $hh$: 74.26 ns, and $kk$: 74.63 ns).
Compared to the PBE lifetimes, the upper limit increase from 50 to 75 ns and low symmetry configurations now have the shortest lifetime.
For the ZPL ($\bar{\boldsymbol{\mu}}^{ee,gg}$) the HSE lifetime ranges from 13 to 23 ns ($kh$: 13.51 ns, $hk$: 18.00 ns, $kk$: 19.06 ns, and $hh$: 22.66 ns).
Compared to the PBE lifetimes, the lower limit decrease from 16 to 13 ns and the order between the configurations is the same.
The main conclusion that the ZPL lifetimes are shorter than the absorption still holds regardless of functional.
Note, the change in polarization and energy between the PBE and HSE functional almost cancel out each other when calculating the radiative lifetime, at least for this defect.
The HSE lifetimes are slightly closer to the experimental lifetimes than the PBE lifetimes.
But as a first step, the PBE functional is accurate enough for the ZPL polarization, intensity, and lifetimes.

\section*{Acknowledgments}
Thanks to Andreas Gällström and Ivan Ivanov for the experimental data. Thanks to Viktor Iv\'ady, Rickard Armiento, Adam Gali and Igor A. Abriksov for helpful discussions. The computations were performed on resources provided by the Swedish National Infrastructure for Computing (SNIC). This work was supported by the Knut and Alice Wallenberg Foundation through Grant KAW 2018.0071 and the Strategic Research Area SeRC.

\bibliographystyle{apsrev4-1}
\bibliography{references}

\end{document}